# Perceived Security of E-Learning Portal

Mohd Faiz Hilmi
School of Distance Education
Universiti Sains Malaysia
mfhilmi@gmail.com

Yanti Mustapha
Department of Business Management
Universiti Teknologi MARA
yantimustapha@gmail.com

*Abstract*—**Information technology has made e-learning possible and available on a large scale. Learning management system (LMS) has been widely used and is accessible through the Internet. However, LMS are exposed to various threats. Proper understanding of the threats is required. Furthermore strategy and best practices countermeasures will ensure a safe learning environment. Therefore, this study looks into the information security aspect of LMS. Specifically, there are two main purposes of this study. First, this study provides a review of information security in e-learning environments and explains the importance of information security. Second, this study looks at how student perceived the security of their e-learning portal. A total of 497 students responded to a survey questionnaires. Frequencies analysis was conducted to show the profile of the respondent. Overall, respondent has strong positive perceptions towards security of their LMS. This study serve as an introduction which help LMS administrator to understand the issues and possibilities related to the safety of LMS.**

## I. INTRODUCTION

The advancement of information technology has changed the education landscape. Various new methods of information delivery have emerged [1, 2]. Information technology has made e-learning possible and available on a large scale particularly via the Internet. E-learning is positioned to experience dramatic growth. Recently, The Custodian of the Two Holy Mosques King Abdullah Ibn Abdulaziz Al Saud, the Premier and the Chairman of Higher Education Council, has approved the establishment of Riyadh-based Saudi Electronic University [3].

Thus, e-learning which relies upon the Internet is open to threats [4] and information security has become a preeminent concern. Therefore, this study provides (a) a review of information security in the e-learning environment and explains security decisions that should be made based on the ten domains of information security established by the International Information Systems Security Certification Consortium and (b) an analysis of how student perceived the security of their e-learning portal. The study approaches perceived security from a broader perspective, which not only includes technical aspects such as confidentiality and authentication also refers to a student's comprehensive sense of security and well-being. Students' perceptions of security can be somewhat different from the real security level on an e-commerce site. Several past researchers focused on various aspect of information security as summarized in Table I.

TABLE I. FOCUS OF SELECTED RESEARCH ON INFORMATION SECURITY OF E-LEARNING

| Author | Focus |
| --- | --- |
| [5] | Security elements in e-learning |
| [6] | Information security collaboration between education providers |
| [7] | Information security rating of e-learning system |
| [8] | Information security governance |
| [9] | Information security policy in higher education institution |
| [10] | Security and privacy issues in e-learning |
| [4] | Countermeasures and information security pillars in -learning environment |
| [11] | Feedback and control rights of online learning participants |

In an e-learning environment, information security risk is likely to arise from the situations outlined below [4].
1. Alteration of material by unauthorized people
2. Bogus course material
3. Submitted assignments copied by unauthorized parties
4. Submitted assignment changed by unauthorized parties
5. Marks changed or deleted
6. Non-authorized access to test papers and test content changed
7. People masquerading as students, write tests on behalf of other students
8. Students obtained unauthorized help during examinations
9. Denial of service attempts against course websites
10. Logon information of lecturers and students could be intercepted and misused.

In order to improve security in an e-learning environment, various authors such as [5], outlined the required security elements in an e-learning environment. [4] identified technical and procedural countermeasures to enhance the security of information on e-learning systems. [7] proposed an information security rating system for e-learning environment. By having such a rating system, the capabilities of e-learning systems could be determined. [9] examined the structure and content of information security policies of several higher education institutions and found that existing policies are not comprehensive enough and did not play an effective role within their institutions. From a similar perspective, [8] examined the e-learning governance practices and identified an e-learning ecosystem governance framework of structures, processes, communications and relational mechanisms, and pedagogies and instructional. Meanwhile, [6] suggested that collaboration between e-learning education providers could enhance information security. According to [10] e-learning system must consider the privacy and security needs of the e-learning participants. Feedback and control rights of online learning participants are also important and must be given a proper attention in an e-earning system design and operation [11].

TABLE II. PURPOSE OF THE TEN DOMAINS [12]

| Domain | Coverage/Purpose |
|---|---|
| **Domain One** – Information Security and Risk Management | Investigates and analyzes the current state of security of information by finding loopholes in the systems then applying the proper amount of counter-measures |
| **Domain Two** – Access Control | Protects information and resources from unauthorized access to the information |
| **Domain Three** – Cryptography | Protects CIA using mathematical means such as cryptography and hashing |
| **Domain Four** – Physical (Environmental) Security | Addresses physical, environmental and procedural risks |
| **Domain Five** – Security Architecture and Design | Protects information models and architectural network methods from unauthorized disclosure, modifications, and destruction |
| **Domain Six** – Business Continuity and Disaster Recovery Planning | Outlines a disaster recovery plan (DRP) that contains procedures to reduce damage during and after a tragic event. Outlines a business continuity plan (BCP) that is a long-term plan to keep the business functioning following a disaster |
| **Domain Seven** – Telecommunication and Network Security | Segregates non-trusted networks using devices, architectures, and protocols to protect the trusted network |
| **Domain Eight** – Application Security | Applies security through the life cycle of the software in use |
| **Domain Nine** – Operations Security | Keeps the organization system running securely by ensuring a secure day-to-day operation |
| **Domain Ten** – Legal, Regulations, Compliance and Investigations | Addresses general computer crime legislation and regulations, investigative measures and techniques |

This study explains each of the ten domains within the context of e-learning systems. Table 3 provides a general overview of the applicability of the ten domains to e-learning systems.

TABLE III. SELECTED E-LEARNING FOCUSES OF THE TEN DOMAINS

| Domain | Selected Focus |
|---|---|
| **Domain One** – Information Security and Risk Management | -Policy, procedures, standards and guidelines of e-learning institutions<br>-Audit framework for e-learning institutions<br>-Awareness and training for staff and students |
| **Domain Two** – Access Control | -Access control to the e-learning system<br>-Intrusion detection and prevention system |
| **Domain Three** – Cryptography | -Security of data transmission |
| **Domain Four** – Physical (Environmental) Security | -Physical security of e-learning institutions<br>-Building access<br>-Information protection and management services |
| **Domain Five** – Security Architecture and Design | -Security framework<br>-Hardware and software design |
| **Domain Six** – Business Continuity and Disaster Recovery Planning | -Availability, uninterrupted access to e-learning system<br>-Assessment, development, implementation and management of continuity planning |
| **Domain Seven** – Telecommunication and Network Security | -Secured transmission of voice, data and multimedia.<br>-Perimeter defense (through firewall or similar mechanism) of the e-learning system |
| **Domain Eight** – Application Security | -Secured e-learning application.<br>-Ensure that the use of open source codes is viruses free |
| **Domain Nine** – Operations Security | -Privilege entity controls of staffs and students accessing the e-learning system<br>-Resource protection<br>-Proper and well documented change control management for any changes, modification or upgrades to the e-learning system |
| **Domain Ten** – Legal, Regulations, Compliance and Investigations | -Understanding of laws and regulations governing the e-learning institution<br>-Handling security incidents for the e-learning system |

*A. Domain One*

The first domain deals with information security and risk management. This domain focuses on the need of having comprehensive policies, procedures, standards and guidelines for information security. Organizations which heavily dependent on information and communication technologies (ICT), such as e-learning institutions, must have a comprehensive information security policy in place [13]. The policy, procedures, standards and guidelines must be comprehensive and not just superficial documents [9]. E-Learning institutions must also have an audit framework, awareness programs and training for staffs and students. Furthermore, collaborative leadership will improve practice of e-learning [14].

### B. Domain Two

This domain focuses on access control of the e-learning systems. Beyond accessibility, quality of service is also a factor that must be considered. Regardless of various statistics that may show significant growth in Internet access, students will only benefits access from location conducive for studying [15]. Access control to e-learning systems must be based on an approved policy of the governing institution. In addition, the e-learning system must also have a mechanism to handle intrusion detection and prevention. In order to safeguard copyrighted contents, proper digital rights management systems and processes are necessary [16].

### C. Domain Three

This domain covers cryptography and the need to ensure that data are only understandable to the intended audiences. Information must be encrypted especially before it is transmitted through public networks. Several existing technologies that can provide the appropriate encryption include encryption algorithms, smartcard technologies and certification schemes [17, 18]. However, whatever technology is selected, it must remain user friendly and non-intrusive to the students [17].

### D. Domain Four

Physical (environmental) security is the basis of this domain. [19] proposed a Security Practitioner's Management Model which consisted of five layers. One of the layers is physical security which refers to the actual physical security of the infrastructure, devices, hardware and software. There must be a building access system that controls the movement of people in and out of any facilities that houses hardware and software. E-learning institutions must also be able to provide information protection and management services of for the e-learning system. If the infrastructure is not supported and protected, the flexibility and benefits of e-learning will be short lived [13].

### E. Domain Five

The security architecture must have a solid security framework. A secured operation has been identified as one of the critical success factor for an E-Learning system [20]. Hardware and software must be selected that contribute to institutional security.

### F. Domain Six

This domain refers to business continuity and disaster recovery planning. As part of the CIA framework, availability is an important aspect of an e-learning system. Students and staff have become more dependent on the system for their learning and teaching. System outages will interrupt student learning. Therefore, continuous availability or uninterrupted access to the e-learning system is paramount to the success of the e-learning operation [21]. Thus, institutions must have a continuity plan that is implemented, monitored and revised on a regular and frequent basis.

### G. Domain Seven

Telecommunication and network security are the focal point of domain seven. . Ease of access to the Internet has been identified as one of the critical success factor for e-learning acceptance [22]. E-learning institutions must ensure secured transmission of voice, data and multimedia between the institution and students. Another important aspect is floor control security which is required especially for synchronized communication activities in the online distance learning environment [23]. In addition, the e-learning system must be protected by a perimeter defense such as a firewall.

### H. Domain Eight

Application security is the centre point of domain eight. The Internet is a not secure means of transmitting data. Web applications must ensure that data are transmitted [24]. If open source codes are used, they should be examined thoroughly to ensure that they are virus free.

### I. Domain Nine

Within the scope of operations security, privilege entity must be in place to control staff and students who access the e-learning system. ICT resources must be protected from unauthorized access. Proper and well-documented change control management must be implemented to ensure that any changes, modifications or upgrades to the e-learning system will not interrupted access to the e-learning system. A secured operation has been identified as one of the critical success factor for an E-Learning system [20].

### J. Domain Ten

Legal, regulations, compliance and investigations are the core of domain ten. Various legal issues such as copyright, fair use and work for hire are currently being examined from an e-learning perspective [25]. All administrators, including deans, deputy deans and managers, must have a good understanding of the laws and regulations governing the e-learning institution. They must ensure that their institution strictly follows all the rules and regulations that have been established. There must also be procedures for handling security incidents. Thus, the institution will be prepared to handle any unexpected issues.

## II. RESEARCH METHODOLOGY

Data for this study are collected from 497 students currently enrolled in a distance learning program. Students in this program use e-learning portal as their main interface to course information, lecture notes, assignments, discussion forums and other related learning materials. The students are also able to view live streaming of course lecture through the e-learning portal. In addition to demographic information, students were asked to state their agreement with five statements related to perceived information security of e-learning portal. The statements are listed in table 4. All five statements were rated based on a 5-point Likert scale anchored from 1 (not at all) to 5 (very much).

TABLE IV. QUESTIONNAIRE ITEMS

| Perceived Security |
|---|
| Q1. I perceive e-learning portal as secure |
| Q2. I perceive the information relating to user and e-learning portal transactions as secure |
| Q3. I do not fear hacker invasions into e-learning portal |
| Q4. I believe the information I provide with e-learning portal will be secured |
| Q5. I am confident that the private information I provide with e-learning portal will be secured |

## III. RESULTS AND ANALYSIS

Students who responded to this study consisted of 201 male and 296 female. Figure 1 presented the gender breakdown of the sample profile. Referring to figure 2, fifty percents of the students are between 25 to 30 years old. Figure 3 shows that about 60% of the respondents earns between RM1000 to RM3000 per month. Majority (62%) of the students are married as presented in figure 4.

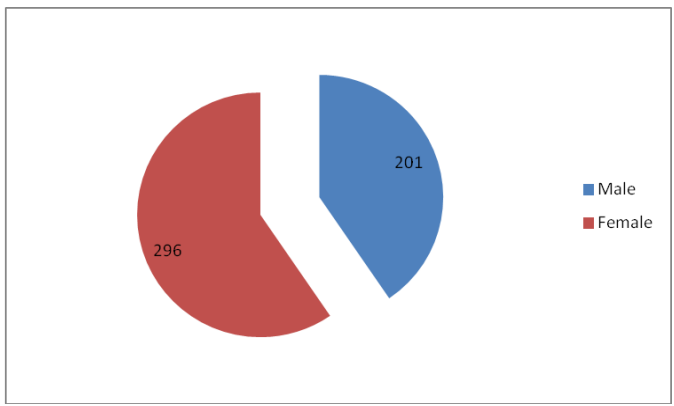

Figure 1. Respondent Profile - Gender

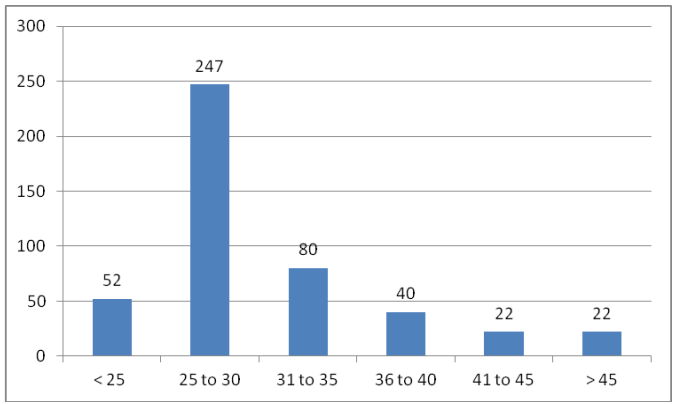

Figure 2. Respondent Profile - Age

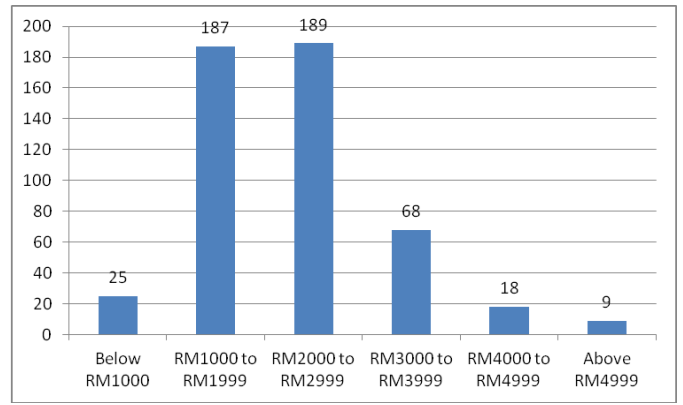

Figure 3. Respondent Profile – Income (RM = Ringgit Malaysia)

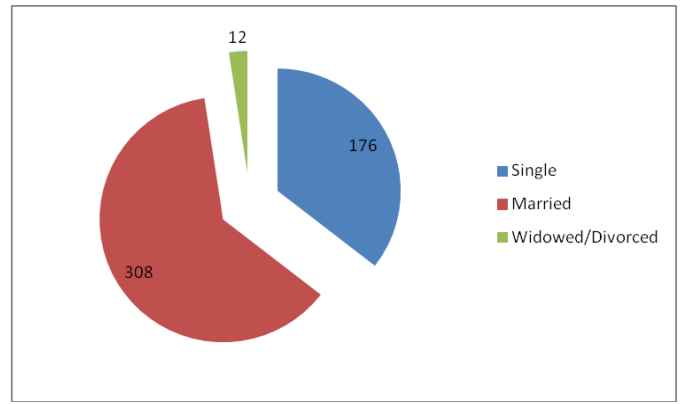

Figure 4. Respondent Profile – Marital Status

Based on a descriptive analysis presented in figure 5, students perceived the e-learning portal as secured. Table 5 and figure 6 listed the actual responses based on five-point Likert scale. Fifty percent of the students perceived the e-learning portal as secured. Forty five percent of the students perceived the information relating to user and e-learning portal transactions as secure. Forty four percent of the students do not fear hacker invasions into e-learning portal and they also believe the information they provide with e-learning portal will be secured. Forty five percent of the students confident that the private information they provide with e-learning portal will be secured.

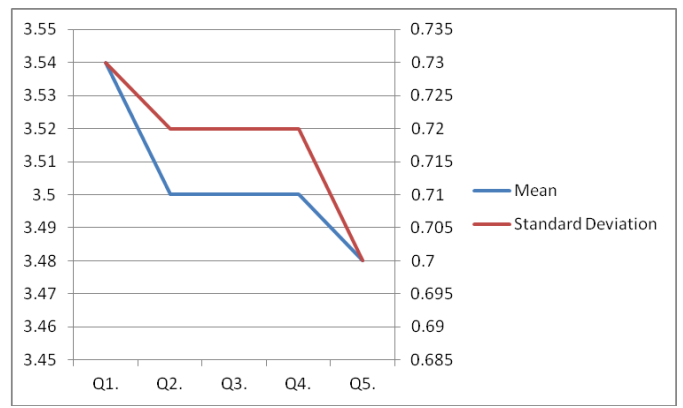

Figure 5. Descriptive Statistics

TABLE V. DETAIL RESPONSES

| Questions | Likert Scale | | | | |
|---|---|---|---|---|---|
| | 1 | 2 | 3 | 4 | 5 |
| Q1. | 3 | 13 | 242 | 188 | 51 |
| Q2. | 3 | 15 | 249 | 184 | 43 |
| Q3. | 1 | 17 | 254 | 178 | 47 |
| Q4. | 1 | 18 | 253 | 175 | 48 |
| Q5. | 1 | 20 | 247 | 191 | 37 |

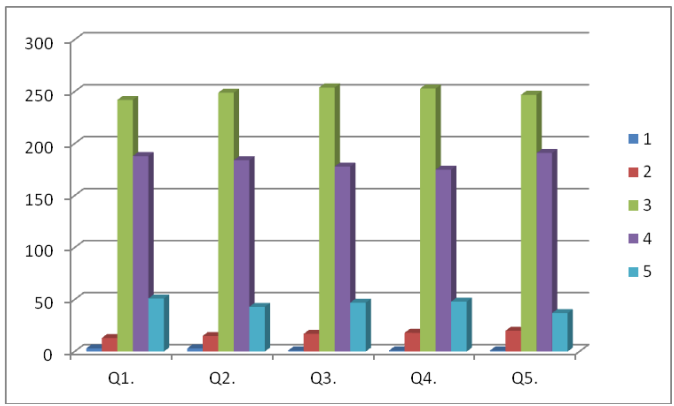

Figure 6. Detail Responses

## IV. DISCUSSION AND CONCLUSION

E-learning portal have arisen as a promising e-learning application for a new type of learning mechanism. But, information technology has also created new threats. It is important for administrator of distance and e-learning to understand how student or distance learners perceived the level of information security of the whole learning system. Perceptions about using the e-learning portal for learning activities will lead to the formation of attitudes that will influence learning behavior.

### A. Review of information security in the e-learning environment

The ten domains of information security established by the International Information Systems Security Certification Consortium provide a comprehensive coverage on all aspects of information technology. LMS administrators should adhere to all the standards and procedures covered by the ten domains to ensure a safe and secured system.

### B. Analysis of how perceived security in e-learning portal affects the students' usage of e-learning portal and learning attitudes

Overall, students exhibit a strong positive perception on the security of their LMS. However, one of the questions was perceived lower than the rest of the questions. "I am confident that the private information I provide with e-learning portal will be secured" reflects the confident that personal data are secured. This study found that student doesn't have a high confident that their personal data within the LMS are secured.

### C. Conclusion

Online distance learning has evolved due to the advancement in information technology. However, information technology has also created new threats. Hackers, viruses and spam are some of the examples. Standards and procedures must be in place to keep online distance learning safe from these threats. One way is to incorporate the information security CBK as part of the online distance learning system. The information security CBK provides comprehensive baseline knowledge and best practices that can be used to improve existing information security architecture and procedures. By closely following the best practices and principles covered by the ten domains, e-learning institutions should be able to provide an e-learning system with high confidentiality, integrity and availability.